# MICROTRIBOLOGICAL PROPERTY OF VERTICALLY ALIGNED CARBON NANOTUBE FILM

Hiroshi Kinoshita*, Ippei Kume, Masahito Tagawa, and Nobuo Ohmae,
*Department of Mechanical Engineering, Kobe University,*
*Rokko-dai 1-1, Nada, Kobe 657-8501 JAPAN*
*TEL/FAX: 078-803-6142, Email: kinohiro@mech.kobe-u.ac.jp*
Viviane Turq, Jean Michel Martin
*Ecole Centrale de Lyon, Laboratoire de Tribologie et Dynamique des Systèmes, UMR CNRS 5513,*
*36, avenue Guy de Collonge, 69134 Ecully Cedex, France*

Microtribological properties of vertically-aligned carbon-nanotube (VACNT) films have been studied [1]. Adhesion forces were obtained by measuring force-displacement curves. Friction experiments were conducted in reciprocating sliding configurations. Effects of tip radius, applied force, scan speed, and relative humidity were investigated. A model of the friction of VACNT film is discussed on the basis of in-situ tribological experiments inside a scanning electron microscope (SEM).

A VACNT film was synthesized by microwave plasma enhanced chemical vapor deposition (CVD) on silicon oxide film. Details of the synthesis method and the characterization of VACNT film were described elsewhere [2]. SEM observations revealed that the thickness of the VACNT film was approximately 6 μm. Multiwalled-type carbon nanotubes with diameters of approximately 20 nm were observed by transmission electron microscopy (TEM). A microtribometer was equipped with a cantilevered probe and an optical lever system to measure adhesion and friction forces. Gold and tungsten tips were prepared by electrochemical etching, and then cleaned in ethanol for 5 min. Probe tips were glued on aluminum cantilevers.

No adhesion force was observed with relative humidity of 0-100 %. Smooth flat friction forces during forward and backward scans in reciprocating sliding were observed, indicating no distinct stick-slip Linear relationship between friction and applied forces were found. From this relationship, friction coefficients of over one were calculated. The superhigh friction coefficient of VACNT film did not depend on the scan speeds, the tip apex radii, the tip materials, and the relative humidity. These friction behaviors obey Amontons-Coulomb's friction law.

It is well known that friction of two clean metal surfaces in a vacuum environment is extremely high due primary to the high adhesion force. The VACNT film had no adhesion force. In-situ tribological experiments inside a SEM were carried out to investigate the friction mechanism of VACNT film. No wear tracks were formed on the contact region of the VACNT film after friction experiments. These facts imply that the VACNT film should have the other friction mechanism which causes extremely high friction.

Wong et al. reported that the bending force of individual carbon nanotubes follows beam theory [3]. Assuming that carbon nanotubes act as cantilever beam and bending of the carbon nanotubes follows beam theory, a simple model of VACNT film friction has been proposed. Figure 1 depicts the model as a tip moves on the surface of a VACNT film. During the friction process, the tip continuously pushes carbon nanotubes with keeping an applied force. At this moment, the bending of the carbon nanotubes gives high repulsive forces, $F_R$, to the tip surface. With respect to the horizontal line $F_R$ can be resolved into a lateral force, $F_L$, and a perpendicular force, $F_P$. The friction force of a VACNT film is computed by the sum of $F_R$ in the conditions of a carbon nanotube radius of 10 nm, length of 6 μm, area density of 300 nanotubes/μm$^2$, Young's modulus of 1.2 TPa, repulsive force directions of 45° with respect to the horizontal line, tip radius of 20 μm, and an applied force of 2.1 μN. Friction force and indentation depth are calculated as 2.1 μN and 1.0 μm, respectively, indicating that the tip does not make contact with the substrate. Thus it is most likely that VACNTs deform elastically to maintain sliding of the tip, and this causes superhigh friction.

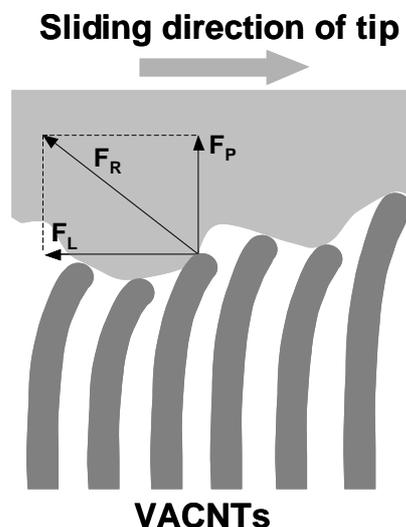

Figure 1 Model showing the mechanical action during friction of VACNT film.

*To whom all correspondence should be addressed.